\documentclass[osajnl,twocolumn,showpacs,superscriptaddress,10pt]{revtex4-1} %% use 11pt for Applied Optics
\usepackage{amsmath,amssymb,graphicx}
\begin{document}

\title{Optimal sub-Poissonian light generation from twin beams by photon-number resolving detectors}

\author{Marco Lamperti}
\affiliation{Dipartimento di Scienza e Alta Tecnologia, Universit\`a degli Studi dell'Insubria, Via Valleggio 11, 22100 Como, Italy}

\author{Alessia Allevi}\email{Corresponding author: alessia.allevi@uninsubria.it}
\affiliation{Dipartimento di Scienza e Alta Tecnologia, Universit\`a degli Studi dell'Insubria, Via Valleggio 11, 22100 Como, Italy} 
\affiliation{CNISM UdR Como, Via Valleggio 11, 22100 Como, Italy}

\author{Maria Bondani}
\affiliation{Istituto di Fotonica e Nanotecnologie, CNR, Via Valleggio 11, 22100 Como, Italy}
\affiliation{CNISM UdR Como, Via Valleggio 11, 22100 Como, Italy}

\author{ Radek Machulka}
\affiliation{RCPTM, Joint Laboratory of Optics of PU and Inst. Phys. AS CR, 17. listopadu 12, 77146 Olomouc, Czech Republic}

\author{V\'{a}clav Mich\'{a}lek}
\affiliation{Institute of Physics AS CR, Joint Laboratory of Optics, 17. listopadu 50a, 77146 Olomouc, Czech Republic}

\author{ Ond\v{r}ej Haderka}
\author{Jan Pe\v{r}ina Jr.}
\affiliation{RCPTM, Joint Laboratory of Optics of PU and Inst. Phys. AS CR, 17. listopadu 12, 77146 Olomouc, Czech Republic}

\begin{abstract} 
We generate nonclassical conditional states by
exploiting the quantum correlations of multi-mode twin-beam states
endowed with a sizeable number of photons. A strong relation
between the sub-shot-noise correlations exhibited by twin beams
and the sub-Poissonian character of the conditional states is
experimentally revealed. It determines optimal conditions for
sub-Poissonian light generation.
\end{abstract}

%\ocis{270.0270, 270.5290, 190.4410, 230.5160}

\maketitle 

\section{Introduction}
During the last decade, the generation of nonclassical states in
the continuous variable domain by means of conditional
measurements has been extensively investigated for many
quantum-optical applications, including quantum information
processing, quantum computing and quantum cryptography
\cite{walmsley05,ralph06,kok07}. In general, conditional
quantum-state preparation schemes benefit from the existence of
correlations between a signal mode and an ancilla system, such as
two output ports of a beam-splitter \cite{sasaki07}, signal and idler modes in
spontaneous parametric down-conversion (SPDC) \cite{Haderka2005a},
cavity mode and atomic levels in cavity QED \cite{haroche01}.
The schemes are based on the fact that, when some observable is measured on the ancilla,
the state of the signal mode is irreversibly modified.\\
In principle, conditional measurements are not directly related to nonclassicality since
also classical conditional states can be prepared ($e.g.$ by
photon subtraction on thermal states or phase-averaged coherent
states \cite{allevi10a,allevi12}). \\
From the experimental point of view, the production of
nonclassical states by means of conditional measurements has been
achieved in the macroscopic regime by starting from a nondegenerate
optical parametric oscillator operated above threshold \cite{laurat03,laurat04},
in the mesoscopic domain by selecting a small specific interval in a very noisy condition \cite{bondani07}, and at very low level by
a number of photon-counting detectors, mostly operated in single-photon regime \cite{kim05,zavatta07,ourj07,taka10,branczyk10} .\\
It is worth noting that the use of 
photon-counting detectors instead of single-photon detectors
offers the possibility not only to enhance the heralding of
single-photon states by suppressing higher photon-number
components \cite{PerinaJr2001,Haderka2004,christ12}, but also to
perform multiple photon-counting operations in order to produce
quasi-Fock states endowed with a number of photons larger than 1.\\
The possibility to extend the experimental results presented so
far to a more mesoscopic photon-number domain, where the states are
more robust with respect to losses, is thus desirable and still
subject to active research. Indeed, the main limitation to achieve
this goal is related to the performances of the available
detectors. For instance, the fiber-loop detectors can work only at
very low light level due to number of single-photon detectors that
constitute their structure \cite{rehacek03,fitch03,Haderka2004},
Silicon photomultipliers are characterized by dark counts and
cross-talk effect \cite{afek09,ramilli10,dovrat12,kala12}, EMCCD
and iCCD cameras are rather noisy because of the gain spreading
\cite{Haderka2005a,blanchet08}, superconductors, such as the
transition-edge sensor (TES) \cite{lita08}
and the superconducting nanowires \cite{goltsman01} must operate at cryogenic temperatures and thus are rather cumbersome.\\
The detectors we used to perform our experimental work are
commercial photon-counting detectors, usually called hybrid
photodetectors (HPD, mod. R10467U-40, Hamamatsu, Japan), whose
main limitation is given by the actual detection efficiency, which
is much lower than $50 \%$ due to the non perfect photon-number
resolution. Nevertheless, in this work we demonstrate that
thanks to the good linearity of such detectors, it is possible to
generate nonclassical conditional states from a multimode twin
beam (TWB) with sizeable numbers of photons. A systematic
study of the quantum properties of the TWB state, given in terms
of the noise reduction factor \cite{bondani07}, is performed to
show that the optimal generation of sub-Poissonian conditional
states strongly depends on the existence of nonclassical
correlations \cite{Perina2007,Perina2009,PerinaJr2013a}, but it is also affected by
several other experimental parameters.

%%%%%%%%%%%%%%%%%%%%%%%%%%%%%%%%%%%%%%%%%%%%%

\section{Multimode TWB in the mesoscopic photon-number regime}
In general, TWB states are intrinsically spectrally multimode
\cite{Perina2005}. We have already demonstrated that a compact
description of multimode TWB is given by
\cite{allevi10b}
\begin{equation}
\label{eq:psiTWB}
|\psi_{\mu}\rangle= \sum_{n=0}^\infty \sqrt{p^\mu_n}
|n^{\otimes}\rangle_s \otimes|n^{\otimes}\rangle_i,
\end{equation}
where $|n^{\otimes}\rangle = \delta(n-\sum_{h=1}^{\mu}n_h)\,\otimes_{k=1}^{\mu}|n_k\rangle$
represents the $n$-photon state coming from $\mu$ equally-populated modes that impinge on the detector and
\begin{equation}
\label{multither}
p^\mu_n= \frac{(n +\mu-1)!}{{n!(\mu - 1)! (N / \mu+1)^{\mu} (\mu/  N +1)^{n}}}
\end{equation}
is the multimode thermal photon-number probability distribution
for $N$ mean-photon number. The TWB state in Eq.~(\ref{eq:psiTWB})
exhibits pairwise correlations, which are preserved even when the state is detected by non-ideal detection
efficiency ($\eta<1$). The quantification of such correlations can be
experimentally obtained by measuring the noise reduction factor
$R=\sigma^2 (m_1-m_2)/ \langle m_1 +m_2 \rangle$, where $m_j$ is
the number of detected photons in the $j$ arm, $ \langle m_1
+m_2 \rangle $ represents the shot-noise level, symbol $\langle \rangle$ denotes the mean value
of the distribution and $\sigma^2()$ indicates the variance. $R$ is a good
nonclassicality criterion because it is possible to demonstrate
that for nonclassical states $1- \eta <R<1$ \cite{bondani07}. When
one of the two parties, say the idler beam, is detected and $m_2$
photons are obtained in the measurement, it is possible to demonstrate that
the conditional state in
the signal arm is characterized by a Fano factor $F=\sigma^2(m)/\langle m \rangle$ that, in the case
of a multimode TWB state \cite{Perina2008a}, reads as follows
\begin{equation}
\label{Ffactor}
F=(1-\eta)\frac{M(m_2+\mu)(M+\eta \mu)}{(M+\mu)[(m_2+\mu)(M+\eta \mu)-\eta \mu (M+\mu)+1]}
\end{equation}
where $\eta$ is the overall detection efficiency that is assumed equal in the two arms and for each of the $\mu$
modes, and $M$ is the mean number of detected photons of the unconditioned state: $M=\eta N$.\\
%%%%%%%%%
\begin{figure}[htbp]
\centerline{\includegraphics[width=1\columnwidth]{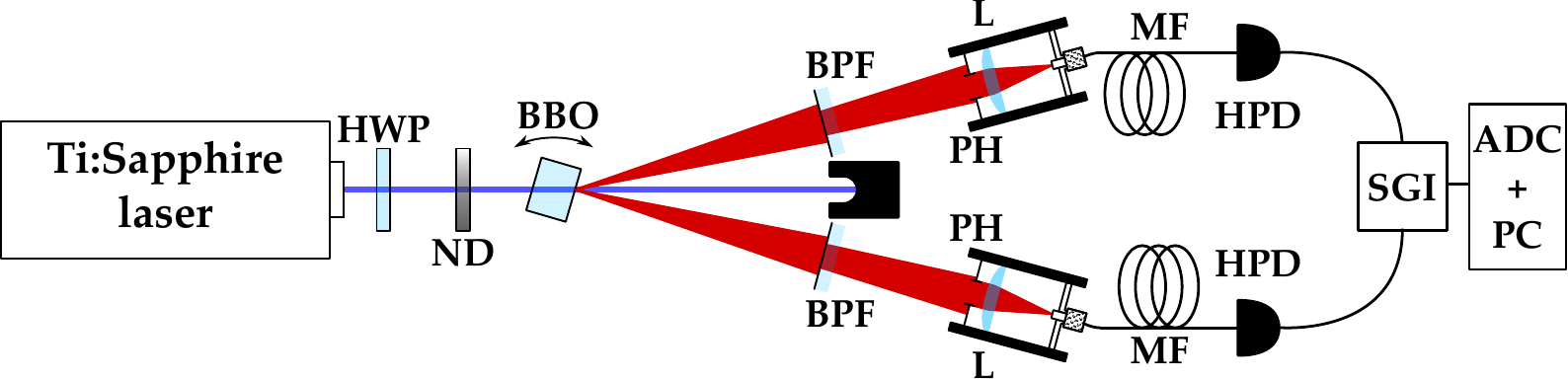}}
\caption{Sketch of the experimental setup. See the text for details.}
\label{setup}
\end{figure}
%%%%%%%%%

As shown in Fig.~\ref{setup}, the SPDC process was obtained by
sending the third-harmonics (at 266 nm) of a cavity-dumped
Kerr-lens mode-locked Ti:Sapphire laser (Mira 900, Coherent Inc.
and PulseSwitch, A.P.E.) to a type-I BBO crystal (8x8x5 mm$^3$,
cut angle $\vartheta_c=48 \deg$). The pulses, whose duration was
set at 144 fs, were delivered at 11 kHz to match the maximum
repetition rate of the detection apparatus. The pump-beam profile
was well approximated by a plane wave, its polarization was
adjusted by means of a half-wave plate (HWP) and its power was
changed by a variable neutral density filter (ND). We collected
two non-collinear frequency-degenerate (at 532 nm) parties of the
TWB state at a distance of 200 mm from the BBO crystal. 
In each arm the light, spectrally filtered by a
bandpass filter (BPF), was selected by an iris with variable
aperture (PH), focused into a multimode fiber (MF, 600-$\mu$m-core
diameter) and delivered to a photon-counting detector. In
particular, we used two hybrid photodetectors, whose declared
quantum efficiency is $\sim 50\%$ in the spectral region we
investigated \cite{bondani09a,bondani09b}.
%%%%%%%%%
\begin{figure}[htb]
\centerline{\includegraphics[width=1\columnwidth]{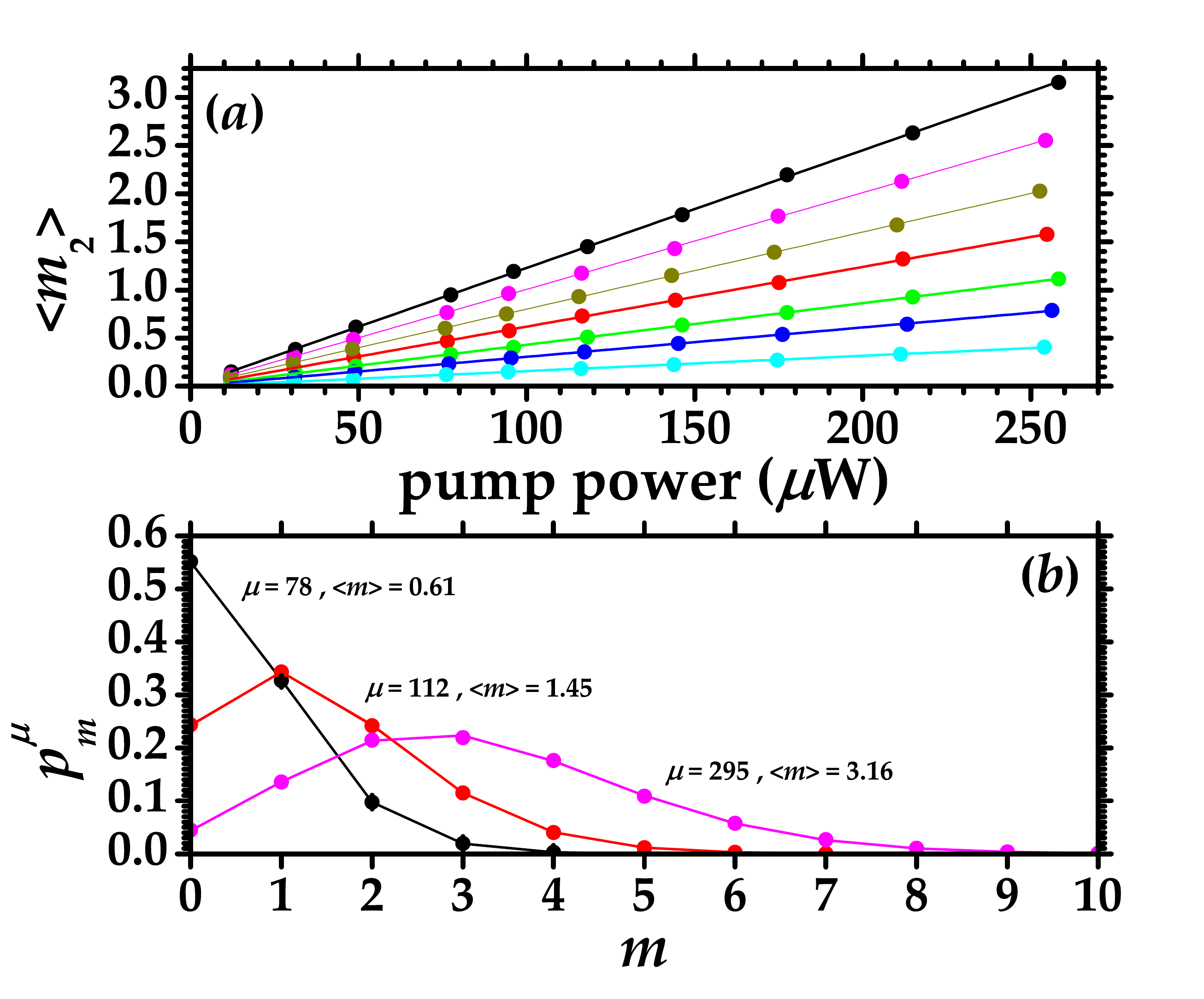}}
\caption{Panel(a): Mean number of detected
photons in the idler arm as a function of the pump mean powers for different iris sizes.
Dots: experimental data, lines: linear fits.
Panel(b): Detected photon-number distributions of some realizations of TWB.
Dots: experimental data, lines: theoretical expectations according to Eq.~(\ref{multither}).}
\label{regime}
\end{figure}
%%%%%%%%%
The TWB was generated in the linear gain regime, as testified by
Fig.~\ref{regime}(a), in which we show the mean number of photons
measured in the idler arm as a function of the pump mean power for
different choices of the iris size (200000 subsequent laser
shots). The fitting curves are indeed straight lines, as it is
expected in the case in which the gain of the SPDC process is
linear. The same figure also shows that our study was performed 
on states containing sizeable numbers of photons (up to 3.2 mean detected
photons), as also proved by some detected-photon distributions
presented in Fig.~\ref{regime}(b).

%%%%%%%%%%%%%%%%%%%%%%%%%%%%%%%%%%%%%%%%%%%%%%%%%%%

\section{Experimental characterization of quantum properties of TWB}
\begin{figure}[htb]
\centerline{\includegraphics[width=1\columnwidth]{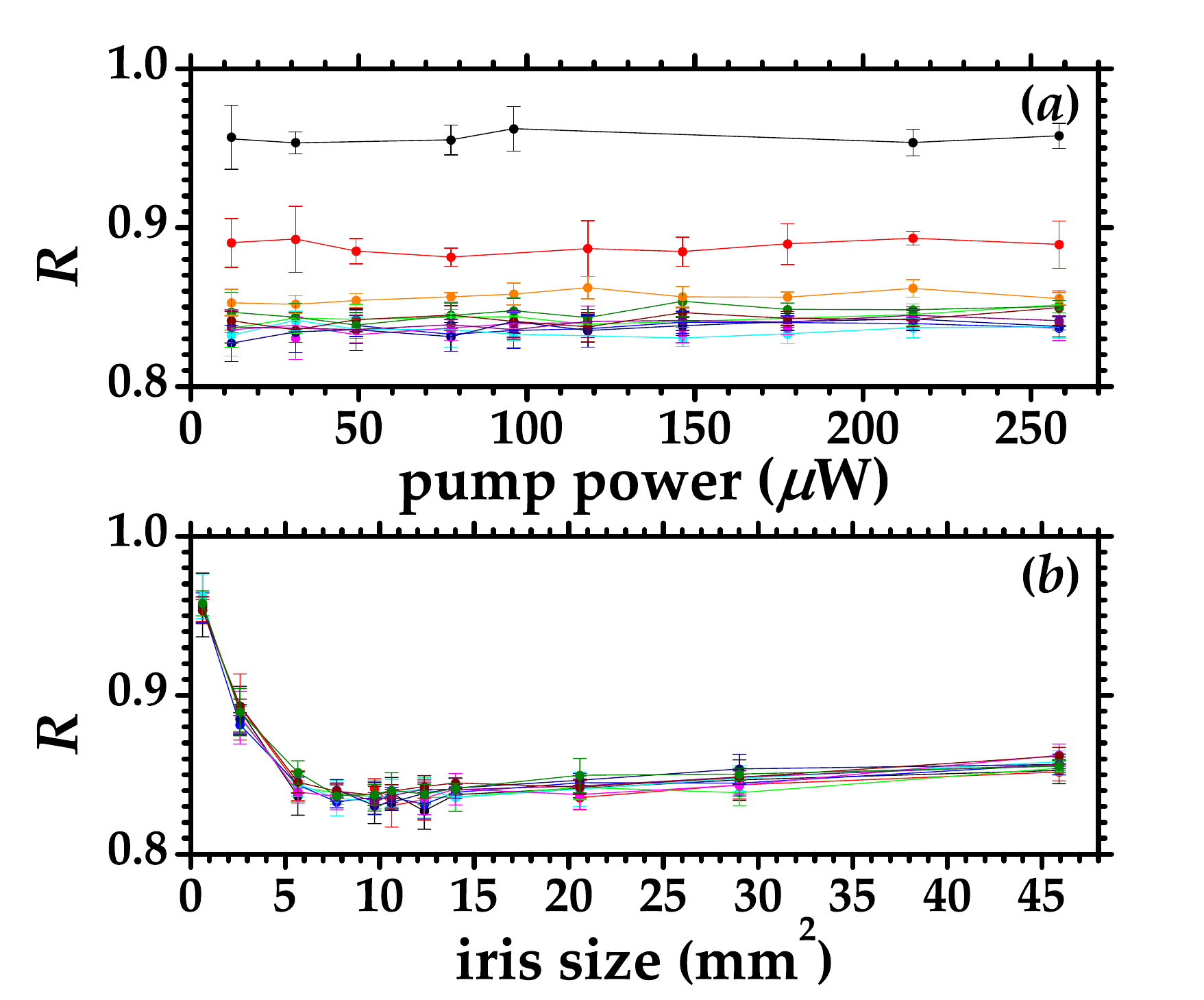}}
\caption{Panel (a): Noise reduction factor $R$ as a function of the
pump mean power for fixed choices (different colours) of the iris
size. Panel (b): $R$ as a function of the iris
size for fixed choices (different colours) of the pump power.
Colour dots: experimental data; coloured lines: theoretical
expectations.}
\label{NRF}
\end{figure}

Quantum properties of the TWB state were investigated by measuring
the pulse-by-pulse sub-shot-noise correlations at different pump
mean powers and for different choices of the collection iris size.
It is important to remark that our results have been obtained in
terms of detected photons by processing the experimental data in a
self-consistent way \cite{bondani09a,bondani09b} without any
$a$-$priori$ calibration of the detection chain and any background
subtraction. In Fig.~\ref{NRF}(a), we plot $R$ as a function of
the pump mean power for different choices of the iris size. It is
almost evident that the noise reduction is independent of the pump
power, which was changed by means of a half-wave plate in order to
keep the beam profile constant. In fact, we have already
demonstrated that changing the pump beam intensity by also varying
the beam size determines a strong variation of $R$
\cite{bondani07}. Moreover, it is important to notice that this
behavior is more evident in the macroscopic regime, in which the
effects of the electronic noise on the one side and of the laser
fluctuations on the other side make the investigation of quantum
properties more challenging. The dependence of the noise reduction
factor on the iris size for fixed choices of the pump mean power
is shown in Fig.~\ref{NRF}(b), in which $R$ exhibits a minimum
as a function of the iris size.
%%%%%%%%%
%%%%%%%%%
The minimum occurrs when the irises are $\sim 3$-mm wide and
coincides with the selection of the largest possible correlated
portions of the twin cones \cite{agafonov10}. When the irises are
smaller, the number of collected photon pairs decreases and the
noise-reduction factor $ R $ increases because of geometrical
filtering inside the correlated areas of twin beams
\cite{Hamar2010}. On the other hand, larger values of iris sizes
exceeding the width of the cone do not substantially increase the
collection of twin components, but the detection of unpaired areas
contributions. In both panels of Fig.~\ref{NRF}, the theoretical
expectations, shown as lines for guiding the eye, are obtained by using the
experimental parameters, determined in a self-consistent way as
described in \cite{allevi12}, in the definition of $R$ given in
\cite{jointdiff} for the case of a single-mode state and here
extended to the multimode case. In particular, the multimode
expression of $R$ reads as
\begin{equation}
R=1-2\eta\frac{\sqrt{\langle m_1\rangle \langle m_2 \rangle}}{\langle m_1\rangle+\langle m_2\rangle}
+\frac{(\langle m_1\rangle \langle m_2 \rangle)^2}{\mu(\langle m_1\rangle +\langle m_2\rangle)},\label{eq:Rteo}
\end{equation}
where $\langle m_1\rangle$ and $\langle m_2\rangle$
are the experimental mean values of detected photons, $\eta$ was calculated by imposing $R=\sigma^2(m_1-m_2)/(\langle m_1\rangle+ \langle m_2\rangle) = 1-\eta$,
that is by assuming that the measured state is actually a twin beam and $\mu$ is the average of the values of the modes in signal and idler
which have been determined by using the first two moments of the detected-photon distribution, namely
$\mu_{i}=\langle m_i\rangle^2/(\sigma^2(m_i)-\langle m_i\rangle)$, $i=1,2$.

%%%%%%%%%%%%%%%%%%%%%%%%%%%%%%%%%%

\section{Sub-Poissonian light generation}
\begin{figure}[htb]
\centerline{\includegraphics[width=1\columnwidth]{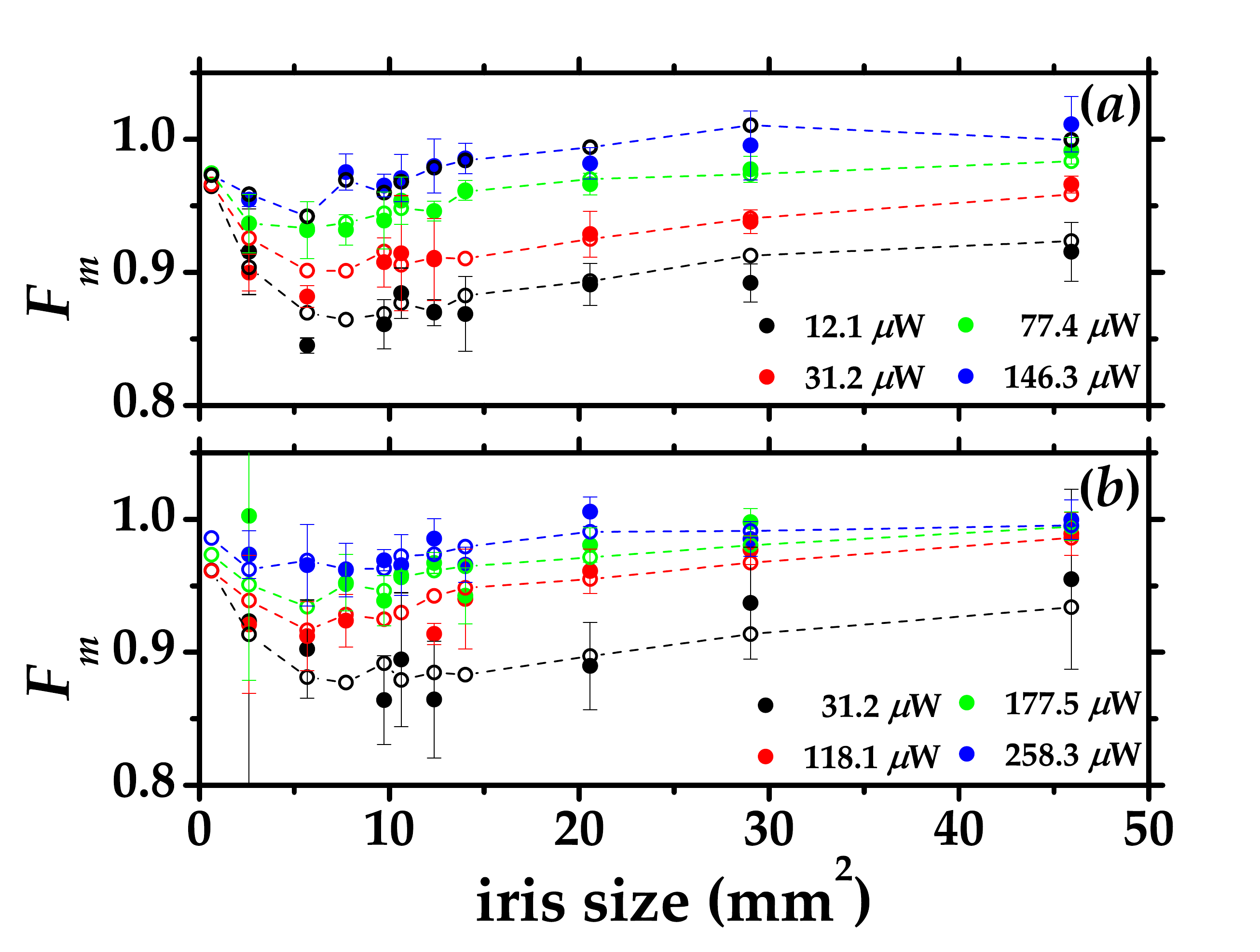}}
\caption{Fano factor $F_m$ of the conditional states obtained on
the signal by detecting $m_2=1$ (panel (a)) and $m_2=2$ (panel (b)) photons on the idler. The
different colours correspond to different pump mean values. Dots:
experimental data, Open symbols + lines: theoretical expectations.}
\label{m=1}
\end{figure}

By exploiting the TWB states characterized above, we performed
conditional measurements by selecting one or two photons in the
idler beam, thus obtaining the conditional states
$\varrho_{m{_2}=1}$ and $\varrho_{m{_2}=2}$ on the signal beam,
respectively. This choice gives us the possibility to show that
our detection apparatus can be used to perform conclusive photon
subtractions by exploiting the photon-counting capability of our
detectors. In particular, we remark that the results obtained in
the case $\varrho_{m{_2}=1}$ are definitely different from those
that can be achieved by employing single-photon detectors operated
in Geiger ON/OFF mode because we do not need to assume that the output states contain 1 photon at most.
The two conditioning operations presented
in this work are useful to investigate the dependence of the
nonclassical nature of the conditional states on the different
experimental parameters involved in their production. The
non-classicality of the conditional states can be quantified by
measuring the Fano factor of detected photons. In general, for any
state detected with Bernoullian probability, the mean value and the variance
of the detected-photon distribution
read as $\langle m \rangle = \eta \langle n \rangle$
and $\sigma^2 (m)= \eta^2 \sigma^2(n)+\eta (1-\eta) \langle n \rangle$, 
respectively, where $\eta$ is the overall detection efficiency. The 
Fano factor for detected photons is thus given by $F_m= \eta F_n+ (1-
\eta)$, where $F_n$ is the Fano factor for photons. If $F_n<1$, the light is
nonclassical and is called sub-Poissonian. Note that the value $1$
for the boundary between classical and nonclassical behavior still
holds for detected photons. It is interesting to notice that the
minimum value of this expression coincides with the minimum value
of noise reduction factor $R = 1-\eta$ in the noiseless limit. In
Fig.~\ref{m=1} we plot $F_m$ for $\varrho_{m{_2}=1}$ (panel (a))
and $\varrho_{m{_2}=2}$ (panel (b)) as a function of the iris size
for four choices of the pump power. We ascribe the discrepancy
between experimental data (coloured dots) and theoretical
expectations (coloured open symbols + lines, which are a guide for
the eye) to possible fluctuations caused by the limited number of
experimental data giving the conditional states. For the same
reason, the data corresponding to the same low pump mean power
($i.e.$ 31.2 $\mu$W) are characterized by bigger error bars in the
case $\varrho_{m{_2}=2}$. Despite this, the minima of $F_m$ as
functions of the iris size coincide with the minimum value of $R$
of the TWB states. We notice that the irregularities in the
theoretical curves are due to the fact that, as in the case of Fig.~\ref{NRF}, 
the theoretical expectations were evaluated at each data point in the very values
of the experimental parameters (the mean number of photons is in
the range 0-3.2, the number of modes varies from 2 to 200, whereas
the overall detection efficiency changes from 0.06 to 0.17) as
calculated from the data.
%%%%%%%%%
%%%%%%%%%
\begin{figure}[htb]
\centerline{\includegraphics[width=1\columnwidth]{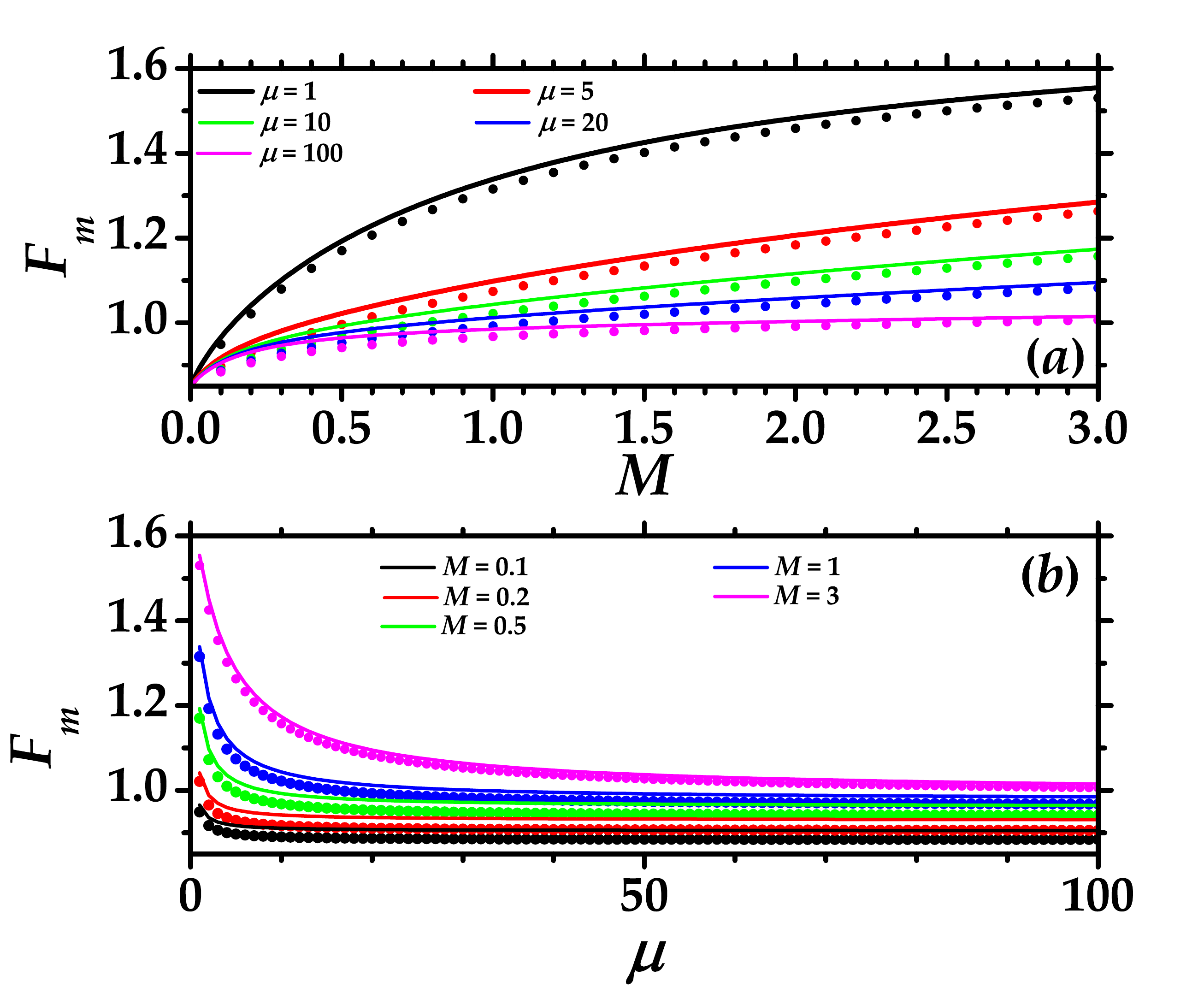}}
\caption{Panel (a): Simulations of $F_m$ as a function of $M$
for different choices of $\mu$. Panel (b): Simulations of $F_m$ as a function of $\mu$
for different choices of $M$. In both panels the quantum efficiency
is $\eta=0.15$ and the conditioning values are $m_2=1$ (solid lines) and $m_2=2$ (dots).}
\label{simul}
\end{figure}
%%%%%%%%%
The experimental results confirm that the amount of nonclassicality in a
twin-beam state represents the critical parameter for
sub-Poissonian-light generation.
Moreover, the largest sub-Poissonianity, that is the lowest value of Fano factor $ F_m $, is obtained for
low pump powers, i.e. when the twin beam is weak. This experimental
conclusion is predicted by theory considering $F_m$
as a function of $M$ for $\eta=0.15$ and different choices of $\mu$ in
the case $m_2=1$ (solid line in Fig.~\ref{simul}(a)) and $m_2=2$ (dots). It is
evident that the lower the mean value of the unconditioned
state, the more sub-Poissonian the conditional state is.
However, we note that the weaker the twin beam the smaller the
post-selection probability in this regime, which imposes
restrictions from the practical point of view.

It is worth noting that also the number of modes $\mu$
influences the achievable values of Fano factor $ F_m $. The larger
the number of modes, the smaller the values of Fano factor
(see Fig.\ref{simul}(b)). In fact, larger values of $\mu$ lead to narrower
photon-pair statistics (transition from thermal to Poissonian
distributions) that improves the post-selection capability of the
applied scheme for non-ideal detection efficiency $\eta$.

%%%%%%%%%%%%%%%%%%%%%%%%%%%%%%%%%%%%%%

\section{Conclusions}
In conclusion, we have experimentally generated sub-Poissonian conditional states
in the low gain regime starting from a multimode TWB state.
By exploiting the linearity of our hybrid photodetectors,
we have confirmed the correspondence of optimal generation conditions
of sub-Poissonian conditional states with those needed for
nonclassical pairwise correlations, even in the case of a limited quantum efficiency.
In principle, similar results can be obtained by choosing larger conditioning values.
Of course, to achieve this goal, larger samples of data are required in order to accumulate enough statistics.
Moreover, the experimental scheme presented in the paper could be exploited for the production
of conditional states optimized to have selected properties,
such as a given mean value or a given amount of sub-Poissonianity.
Finally, in a previous paper \cite{allevi10b} we have already demonstrated that conditioning operations generate
conditional states that are also non-Gaussian. This property, together with nonclassicality, is crucial for the
realization of some quantum information protocols, such as entanglement distillation.
However, according to the simulations presented in Ref.~\cite{allevi10b}, the behavior of
non-Gaussianity as a function of the different experimental parameters
is not the same as that exhibited by the Fano factor. Work is in progress to
find the best compromise between sub-Poissonianity and non-Gaussianity for the
exploitation of the states in real protocols.

\section{Acknowledgments}
The research leading to these results has been supported by MIUR
(FIRB ÒLiCHISÓ - RBFR10YQ3H). Support by projects P205/12/0382 of
GA \v{C}R and projects CZ.1.05/2.1.00/03.0058 and
CZ.1.07/2.3.00/20.0058 of M\v{S}MT \v{C}R are acknowledged.

\end{document}